\documentclass[sn-mathphys-num]{sn-jnl}

\usepackage{setspace}
\usepackage{dsfont}
\DeclareUnicodeCharacter{2005}{\hspace{0.25em}}
\usepackage{comment}
\usepackage{makecell}
\usepackage{graphicx}%
\usepackage{multirow}%
\usepackage{amsmath,amssymb,amsfonts}%
\usepackage{amsthm}%
\usepackage{mathrsfs}%
\usepackage[title]{appendix}%
\usepackage{xcolor}%
\usepackage{tabularx}
\usepackage{textcomp}%
\usepackage{soul}
\usepackage{manyfoot}%
\usepackage{booktabs}%
\usepackage{algorithm}%
\usepackage{algorithmicx}%
\usepackage{algpseudocode}%
\usepackage{listings}%
\usepackage{footnote}
\usepackage{array}
\usepackage[normalem]{ulem}
\makesavenoteenv{table}
\makesavenoteenv{tabular}

\begin{document}

\title[Article Title]{Explicit attacks on differential phase shift quantum key distribution}

\author[1]{\fnm{Valliamai} \sur{Ramanathan}}

\author[1]{\fnm{Anil} \sur{Prabhakar}}

\author[2]{\fnm{Prabha} \sur{Mandayam}}

\affil[1]{\orgdiv{Department Of Electrical Engineering}, \orgname{Indian Institute of Technology, Madras}, \orgaddress{\street{}, \city{Chennai}, \postcode{600036}, \country{India}}}

\affil[2]{\orgdiv{Department Of Physics}, \orgname{Indian Institute of Technology, Madras}, \orgaddress{\street{}, \city{Chennai}, \postcode{600036}, \country{India}}}

\abstract{
 {In the well-established framework of quantum key distribution (QKD), differential phase shift (DPS) protocols have known information-theoretic security bounds defining tolerable error rates under optimal adversaries. In this work, we revisit the security of 3- and \(n\)-pulse DPS QKD by explicitly analyzing two specific, physically implementable individual attacks: minimum error discrimination (MED) and quantum cloning. Using semidefinite programming, we characterize these attacks in detail and quantify their induced quantum bit error rates (QBER) and resulting secure key rates under realistic system assumptions. The critical QBER thresholds for these attacks are approximately 20\%, significantly higher than the theoretical lower bounds of 6\% for individual and 4\% for coherent attacks, indicating these are suboptimal adversarial strategies. This study primarily serves as a practical exercise to benchmark known attacks, providing explicit measures that aid experimental validation, protocol calibration, and risk assessment within current technological capabilities. Additionally, we explore finite-size effects and the effectiveness of phase randomization in weak coherent source-based protocols as protection against unambiguous state discrimination attacks.
}

}
\keywords{Quantum Key Distribution, Differential Phase, Minimum error discrimination, Cloning.}



\maketitle
\doublespacing

\section{Introduction}\protect
Quantum Key Distribution (QKD) offers the promise of secure communication over public networks, ideally, with \emph{unconditional security}~\cite{bb84, Bennett1992, ekert1991quantum}. 

The security of QKD relies on fundamental properties of quantum systems including monogamy of entanglement, the no-cloning principle, and the uncertainty principle~\cite{shorpreskill}. Quantitatively, the security of a specific QKD implementation is characterised by its secure key rate \cite{Renner2008,portmann2021security}. Both asymptotic and finite key security analyses have been presented for the standard QKD protocols in the literature~\cite{scarani2008quantum, HayashiFinite,sheridan2010finite,cai2009finite}.

In this paper, we focus on the specific class of differential phase shift (DPS) QKD protocols~\cite{Inoue, InterceptResend2017, DPSInoue}. This class of distributed phase-reference protocols is known for its simplicity and efficient key generation. DPS QKD does not require basis reconciliation as in the case of the BB84 protocol and thus every bit that is detected contributes to the key. The ease of implementation of this protocol makes it of practical interest \cite{shaw2020equivalence,DPSSecurity}. Furthermore, the 3-pulse DPS can be implemented on an integrated photonic
circuit. There are several variations of this protocol including round-robin DPS~\cite{Sasaki2014RRDPS}, the small-number-random DPS protocol \cite{SNRDPS2017}, and measurement-device-independent DPS  (MDI-DPS)~\cite{Ranu_2021}.

While the DPS protocol was originally proven to be secure against basic individual attacks such as the intercept-resend and beam splitter attacks, it was subsequently shown to be secure against more general individual attacks ~\cite{IndividualAttackSecurity}. Unconditional security was then proved under general coherent attacks for single photon DPS QKD \cite{UnconditionalSecurity_PRL}. In proving unconditional security, the eavesdropper is considered to be capable of entangling ancillary systems to blocks of pulses and performing operations that are dependent on previous measurement results. Specific collective attacks on weak coherent-based implementation have also been studied in the past~\cite{collecticeBS}. More recently, there have been formal proofs of security for the weak-coherent-state based DPS-QKD \cite{Akhiro_2023_DPSSproof,sandfuchs2023security,Mizutani2019} protocol.

In protocols like BB84, theoretical limits on key generation are well known, with a quantum bit error rate (QBER) threshold of around 11\%~\cite{shorpreskill} that is achieved by phase-covariant cloning attacks~\cite{PRA_phasecovariant}. We note that the critical QBER for a practical BB84 implementation with weak coherent states is generally lower than the idealized 11\%, depending on factors such as the number of decoy intensity settings, error correction efficiency, and detector characteristics. {Critical QBER is defined as the error rate at which the secure key rate becomes zero. It is obtained by setting the secure key rate expression to zero and solving for the corresponding QBER.} Going beyond BB84, DPS QKD implementations are also now commercially available~\cite{Gupta23}. This motivated us to construct and analyse explicit attacks on DPS QKD, to evaluate their effectiveness relative to theoretical bounds. Our findings reveal that these attacks remain suboptimal, producing higher QBERs than known security limits, highlighting a substantial gap between implementable adversarial strategies and information-theoretic security guarantees. This gap offers insight into the practical noise tolerances achievable in real-world DPS QKD systems and serves as a basis for further exploration of adversarial capabilities and system robustness.

 {In practical quantum key distribution (QKD) implementations, detector technology plays a central role in determining system performance. Superconducting nanowire single-photon detectors (SNSPDs) provide near-ideal characteristics, including ultra-low dark count rates, high detection efficiencies, and low timing jitter, making them the preferred choice for high-performance and long-distance QKD systems . However, their requirement for cryogenic cooling limits their scalability and widespread deployment. In contrast, most field-deployable QKD systems \cite{shaw2020equivalence} rely on InGaAs avalanche photodiodes (APDs), which are compact and cost-effective but introduce higher noise due to dark counts, afterpulsing, and dead-time effects, leading to increased quantum bit error rates (QBER) . As a result, practical QKD systems often operate closer to the security threshold compared to ideal laboratory conditions. In this context, the analysis of individual attack strategies provides a useful benchmark to evaluate the robustness of QKD protocols under realistic conditions and to assess whether a non-zero secure key rate can be maintained in the presence of experimentally relevant imperfections.}

In this work, we study two specific kinds of individual attacks, the minimum error discrimination (MED) attack and the cloning attack. The effects of these attacks on the secure key rate of the DPS QKD protocol are yet to be quantified. 
Since ideal $n\geq3$-pulse DPS states are linearly dependent, they do not permit unambiguous state discrimination, which requires linear independence~\cite{CHEFLES1998339}; hence, we focus on minimum error discrimination instead.
In past work, an intercept and resend (IR) attack has been considered wherein the eavesdropper (Eve) uses the same setup as the receiver (Bob) \cite{InterceptResend2017}, we will refer to this attack as IR henceforth. In the current work, we consider the scenario where Eve does minimum error discrimination (MED) of the states after intercepting them. This constitutes the first part of the study. MED is based on a detection theory \cite{Helstrom1969} and has been studied in the past in the context of other phase-based protocols such as the coherent one-way protocol~\cite{MinErrorAttack}.

The second part of our work analyzes the security of the DPS QKD protocol against a cloning attack. The cloning attack has been previously studied for QKD protocols such as BB84 and the six-state protocol~ \cite{CloningOnBB84, PhaseCovarintClonnerBB84, CloningAPS2004}, and has also been used as a counterfeiting procedure for quantum money \cite{molina2012optimal}. The problem of finding the \emph{optimal} quantum cloner is a convex optimization problem and can be formulated as a semidefinite program. In our work, we assume that the eavesdropper uses a symmetric quantum cloning machine that maximizes the cloning fidelity for the $3$-pulse DPS QKD protocol. 

The rest of the paper is organized as follows. In Sec.~\ref{sec:Protocol} we briefly review the DPS QKD protocol. Sec.~ \ref{sec:MEDAttack} describes the MED of quantum states and formulates the MED attack for the $3$-pulse and $n$-pulse DPS protocols. In Sec.~\ref{sec:cloningattack} the description of a cloning attack against $3$-pulse DPS is provided, along with the error rates introduced. In Sec.~\ref{sec:unitarycloning} we deviate from the map-based optimal cloning and study how a unitary cloning operation utilising only one ancillary system performs as compared to the optimal map-based cloning. In Sec.~\ref{sec:WCS} we look at the weak coherent state (WCS) state-based DPS QKD protocol and study the effect of the USD attack on the same. Finally, we quantify the usefulness of phase randomization as a measure against the USD attack.

\section{\label{sec:Protocol} Preliminaries}

\begin{figure*}[ht!]
    \centering
    \includegraphics[width=\columnwidth]{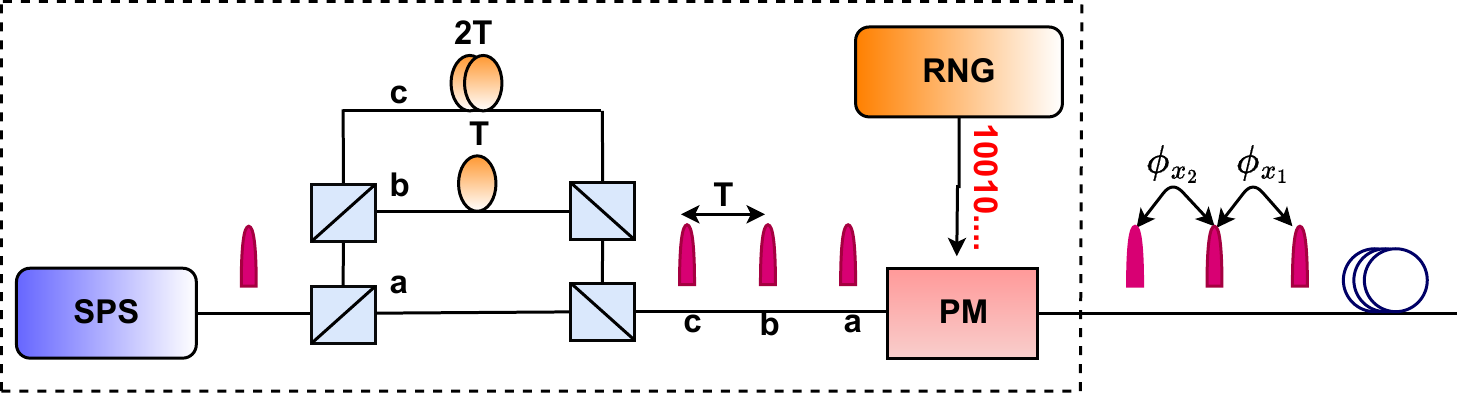}
    \caption{The $3$-pulse DPS QKD protocol. SPS - Single photon source, PM - Phase Modulator, $\phi_{x_1}$ and $\phi_{x_2}$ are the phases introduced between the pulses by the phase modulator, RNG - Random number generator. ${\phi_{x_1}, \phi_{x_2} \in \{0,\pi\}}$. } 
    \label{3PulseDPS}
\end{figure*}

We begin with a brief review of the $n$-pulse DPS QKD protocol~ \cite{Inoue,InterceptResend2017}, schematically shown in 
Fig.~\ref{3PulseDPS} for the $n=3$ case. We note that this schematic can be easily extended to $n$ pulses by simply introducing more delay lines at the source, or using an equivalent time-bin phase modulation~\cite{shaw2020equivalence}. The sender (Alice) prepares a photon in an equal superposition of $n \geq 3$ pulses and encodes her bit values $\{0,1\}$ via the relative phases $\{0,\pi\}$. The phase on the first pulse is considered as the global phase, thus allowing for $n-1$ bits of information to be encoded as relative phases of consecutive pulses. The states sent by Alice in the $n=3$ case can be written as, 
\begin{equation}
    |\psi\rangle_{\rm \pm,\pm}= \dfrac{1}{\sqrt{3}} (|1\rangle_1 | 0\rangle_2 |0\rangle_3 \pm |0\rangle_1 |1\rangle_2 |0\rangle_3 \pm |0\rangle_1 |0\rangle_2 |1\rangle_3).
    \label{eq:eq1}
\end{equation}
Here $|i\rangle_j$ refers to $i$ photons in the $j^{\text{th}}$ path. Or equivalently, a specific DPS state can be written as,
\begin{equation}
    |\psi\rangle_{+,+} = \frac{1}{\sqrt{3}}\left[a_{1}^\dagger + a_{2}^\dagger + a_{3}^\dagger\right]|0\rangle
\end{equation}
where $a_i^\dagger$ refers to the creation operator at the $i^{\text{th}}$ path. The set of states sent by Alice in the $3$-pulse DPS protocol belongs to a three-dimensional Hilbert space spanned by the basis states,
\begin{align*}
    |1\rangle_1 | 0\rangle_2 |0\rangle_3 \triangleq |1\rangle\\
    |0\rangle_1 |1\rangle_2 |0\rangle_3 \triangleq |2\rangle \\
    |0\rangle_1 |0\rangle_2 |1\rangle_3 \triangleq |3\rangle.
\end{align*}
Thus, the signal states sent by Alice can be written in vector form as, 
$$\begin{array}{ll}
|\psi\rangle_{+,+} = (1/\sqrt{3})[1,1,1]^T,\\
|\psi\rangle_{+,-}= (1/\sqrt{3})[1,1,-1]^T,\\ 
|\psi\rangle_{-,+} = (1/\sqrt{3})[1,-1,1]^T,\\  
|\psi\rangle_{-,-} = (1/\sqrt{3})[1,-1,-1]^T. 
\end{array}$$ 

The receiver (Bob) uses an unbalanced Mach Zehnder interferometer (MZI) for decoding, detects the relative phase between the pulses, and hence obtains the bit value sent by Alice. On Bob's side, the detection at the first and the last time slots are random and are not used in generating the shared key. Detection in any of the other $n-1$ time slots gives the key bits, wherein the bit-value is assigned based on which detector (constructive or destructive) clicks. In the ideal case, when a single photon is sent in a superposition of $n$ pulses, the sifted key rate is obtained as,
\begin{equation}
    R_{\text{sifted}}^{(n)} = \frac{n-1}{n}.
\end{equation}
The security of this DPS protocol is based on the fact that the states in Eq.~\eqref{eq:eq1} are non-orthogonal and hence cannot be distinguished perfectly.  For the $n=3$ case, we note that the sifting factor is $2/3$ and is higher than BB84 where the sifting factor is $1/2$.

\subsection{Secure key rate }
 The security of any QKD implementation can be quantified via its secure key rate. Assuming that the eavesdropper is restricted to only individual attacks, the secure key rate can be calculated for the DPS QKD protocol as \cite{DPSSecurity},
\begin{equation}
R_{\text{sk}}= R_{\rm{sifted}}^{(n)}p_{\text {click }}\left[\tau-f(e_b)\left[h(e_b)\right]\right]\label{keyrateExp}.
\end{equation}

\noindent Here,  
$p_{\text {click }}$ is the probability of Bob's detection taking into account detector efficiencies, $e_b$ is the bit error rate, and $f(e_b)$ characterizes the performance of the error correction algorithm. The parameter $\tau$ represents the shrinking factor due to privacy amplification and is calculated from the average collision probability, and, $h(e_{b})$ is the binary Shannon entropy \cite{DPSSecurity}, defined as,
\begin{equation}h(e_b)=-[e_b \log _{2} e_b+(1-e_b) \log _{2}(1-e_b)].\end{equation} 
The key rate given above is in the asymptotic limit. 
 
Based on Shannon's noiseless coding theorem, the minimum number of bits that must be exchanged publicly for error correction equals the binary entropy of the bit error rate $e_b$. The bit error rate $e_b$ has contributions from both imperfect state preparation as well as dark counts. 
 
The dark counts are uncorrelated to Alice's key and thus have a 50 percent chance of being wrong. The bit error rate is thus \begin{equation}e_b=\dfrac{0.5 p_{\text{dark}}+ B\cdot p_{\text{signal}}}{p_{\text{click}}},\label{eq:eb}\end{equation} 
where $B$ is the baseline system error rate quantifying the imperfect state preparation, $p_{\rm dark}$ is dark count probability, $p_{\rm signal}$ is the probability of getting a click due to a photon given channel loss, distance and detector efficiency, and $p_{\rm click} = p_{\rm dark} + p_{\rm signal}$. All the system's errors are usually given as an advantage to Eve and thus the role of privacy amplification is to deduce the shrinking factor $\tau$ by which the key rate has to shrink so we can estimate a bound on the information leaked to Eve. The shrinking factor $\tau$ is a function of the average collision probability $p_c$, \begin{equation} {\tau = - \log_{2} p_c.
}\end{equation} 

The average collision probability quantifies Eve's mutual information with Alice and Bob and can be evaluated in terms of the individual collision probability of each bit, denoted as $p_\text{co}$. In fact, for individual attacks, the overall collision probability is simply the product of the individual collision probabilities. Specifically, $p_\text{co}$ is set to $1$ for the bits that are detected correctly by Eve and $1/2$ for the bits unknown to Eve \cite{DPSSecurity}. In general, the average collision probability is given as \cite{IndividualAttackSecurity} \begin{equation}p_{\text{co}}=\sum_{x,z}p^2(X_i=x|Z_i=z)p(Z_i=z),\label{colleq}\end{equation}  where $X$ and $Z$ are Alice's and Eve's key bit string and Eve's information respectively. In \cite{IndividualAttackSecurity}, the most general individual attack is considered and a bound on the collision probability for each bit is found to be 
\begin{equation}
    p_{\text{co}}\leq 1-e_b^2- {(1-6e_b)^{2}}/2.
\end{equation}
 Using this value in Eq.~\eqref{keyrateExp} gives a lower bound on the secure key rate obtained against individual attacks for DPS QKD \cite{IndividualAttackSecurity},
\begin{equation}
     {R_{\rm ind} = -p_{\rm click}\left[- \log_{2} p_{\rm co} + f(e_b)h(e_b)\right]}.\label{low_ind}
\end{equation}
Here the first term, $-\log_{2} p_{\rm co}$ is the shrinking factor $(\tau_{\rm ~low})$. 

 {Apart from the shrinking factor, we also use the critical quantum bit error rate (QBER) to evaluate the effectiveness of different attacks. This is the error rate at which the secure key rate becomes zero. It is obtained by setting the key rate expression in Eq.~\eqref{low_ind} to zero and solving for the corresponding QBER. Beyond this value, no secure key can be generated, and the protocol must be aborted. From Eq.~\eqref{low_ind}, the critical QBER is found to be $6\%$.}

The unconditional security proof in  \cite{UnconditionalSecurity_PRL} gives a bound on the phase error rate by assuming a general coherent attack and a lower bound on the secure key rate is given as
\begin{equation}
    R_{\text{sk}} \geq R_{\text{sifted}} \left[1-h(e_b)- h((3+\sqrt{5})e_b)\right]. \label{low_coh}
\end{equation}
The critical error rate when considering a general coherent attack is $4~\%$.  {It is obtained by setting the key rate expression in Eq.~\eqref{low_coh} to zero and solving for the corresponding QBER.} 

In our work, we consider an adversary who applies an i.i.d based on MED and cloning. We assume that Eve does not hold any information intercepted in the quantum memory.
We now proceed to discuss the different individual attacks in detail and evaluate the corresponding secure key rates and critical QBERs.

\section{\label{sec:MEDAttack}Minimum error discrimination attack}


 In the minimum error discrimination (MED) attack, the eavesdropper (Eve) aims to identify the optimal quantum measurements required to discriminate the signals transferred through a channel during QKD. The minimum error condition is enforced by maximizing the success probability of the state discrimination. 
Minimum error discrimination can be described mathematically as follows.

Suppose Alice prepares a set of $N$ states  $\{|\psi_1\rangle, |\psi_2\rangle,\ldots,|\psi_N\rangle\}$ which are non-orthogonal and linearly dependent. She chooses the state $|\psi_i\rangle$  with probability $p(i)$ and sends it to Bob. Eve uses a positive operator valued measure (POVM) comprising a set of $n$ positive operators denoted as $\{P_1, \ldots, P_N\} $ such that outcome $i$ associated with measurement operator $P_i$ corresponds to state ${|\psi_i\rangle}$. 
Let $\{\rho_{i} = |\psi_{i}\rangle\langle\psi_{i}|\}$ denote the density operators corresponding to Alice's signal states. Then, the average success probability corresponding to Eve's POVM is $\sum_i p(i) \textrm{Tr} [\rho_i P_i]$. The problem of minimum error discrimination is to maximize this success probability and following \cite{SDPResource}, this can be recast as a semidefinite program (SDP), as described below.

\begin{eqnarray}
\text{maximize}& \sum_{i =1}^{N} \langle \sigma_i, P_i \rangle \nonumber \\
\text{subject to}\\ & P_i \geq 0, \nonumber\\
& \sum_{i=1}^{N} P_i = I. \nonumber
\end{eqnarray}
 
 \noindent where $\sigma_i = p(i) \rho_i$ and $\langle \sigma_i, P_i \rangle = \textrm{Tr} [\sigma_{i} P_i]$ denotes the Hilbert-Schmidt inner product between the positive operators $\sigma_{i}$ and $P_{i}$. We can therefore use an SDP solver to find the optimal success probabilities and the corresponding POVM for minimum error discrimination of a set of quantum states.


\subsection{MED of the ideal $3$-pulse DPS QKD states }
 The ideal single-photon $3$-pulse DPS QKD protocol shown in Fig.\ref{3PulseDPS} encodes the logical bits as relative phases. The states sent by Alice can be written as in Eq.~\eqref{eq:eq1}. We assume that Alice sends the four states with equal probability, that is, $\frac{1}{4}$. Thus to do a minimum error discrimination of these four states, Eve tries to identify the optimal POVM with four elements $\{P_1, P_2, P_3, P_4\}$, where each POVM element correctly identifies one state.
The SDP corresponding to minimum error discrimination of the $3$-pulse DPS states is therefore written as,

\begin{eqnarray}
    \text{ maximize:} & \sum_{i} \frac{1}{4}\langle \rho_i, P_i\rangle \nonumber\\
  \text {subject to:}\\  &\sum_iP_i = I, \nonumber \\ 
& P_i \geq 0. \nonumber
\end{eqnarray}\label{eq:sdp_main}


\noindent where $\rho_i$ is the density matrix corresponding to the DPS states.

\begin{figure}[h]
    \centering
    \includegraphics[width=0.5\linewidth]{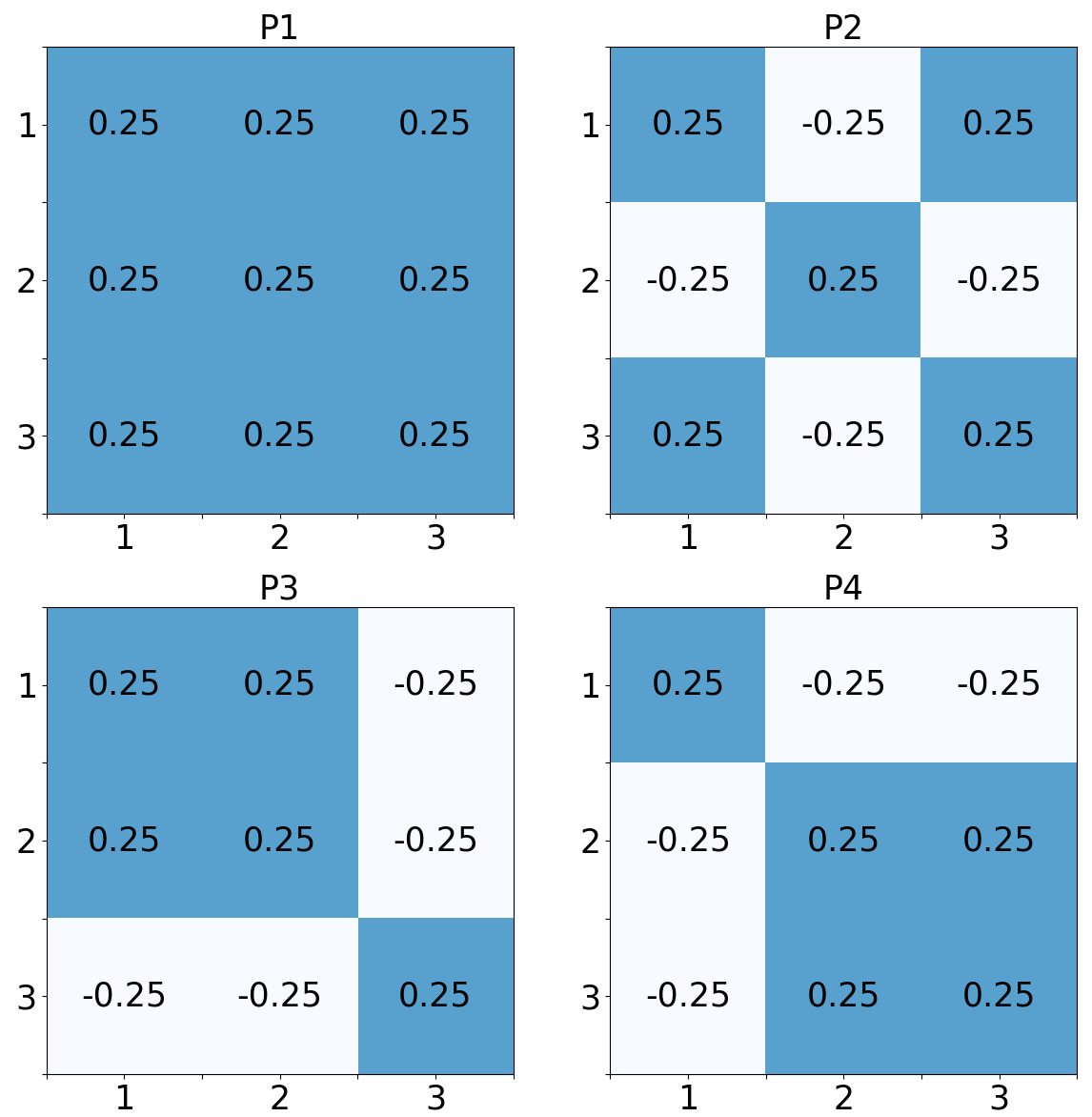}
    \caption{The POVM elements for MED of the four states in the $n=3$ pulse DPS QKD written in the computational basis.}
    \label{fig:MED_POVM}
\end{figure}

We solve this SDP using the cvx solver \cite{cvx,gb08} and summarize the detection probabilities in Table \ref{Table2}. We see that each state has a 75\% chance of being correctly identified by Eve. The positive operators that constitute this optimal POVM are given in Fig~\ref{fig:MED_POVM}. 

 \begin{table}
   \caption{Optimal measurement probabilities for $3$-pulse DPS states. $|\psi\rangle$'s represent the DPS state and $\{P_i\}$ the POVM elements}
\label{Table2}

 \begin{tabular}{ccccc}
 \toprule
     & $P_1$& $P_2$&$P_3$ &$P_4 $ \\
    \midrule
     $|\psi\rangle_{+,+}$&$0.75$& $0.083$ & 0.083 & 0.083 \\
     $|\psi\rangle_{+,-} $&0.083& $0.75$ & 0.083& 0.083 \\
     $|\psi\rangle_{-,+}$&0.083&0.083&0.75&0.083\\
     $|\psi\rangle_{-,-}$&0.083 & 0.083 & 0.083& 0.75  \\
     \botrule
  \end{tabular}
     \end{table}

\subsection{Intercept and resend attack with MED} 

 The optimal MED for the $3$-pulse DPS states shows that Eve has a $75$\% probability of correctly identifying a given state. 
We had previously considered an intercept and resend attack where Eve uses the same apparatus as Bob, and introduced errors in $33\%$ of the intercepted bits \cite{InterceptResend2017}. But in an intercept attack using MED, Eve introduces only a 25\% error. We now compare the key rate in the presence of these two attacks. \par
To find the secure key rate in the presence of a MED attack, one has to find the collision probability and the corresponding shrinking factor. The collision probability is defined as in Eq.~(\ref{colleq}), with $P(X|Z)$ being the probability of guessing Alice's string given Eve's information Z. In the case of a MED attack, Eve's information corresponds to which of the POVMs clicked. Therefore $Z \in \{P_1, P_2, P_3, P_4\}$. Alice's bit value is either 0 or 1 that is $X \in \{0,1\}$, which can, in turn, be detected from the phase difference between the first and the second time bin or the second and third, both the cases being equally probable.
 Now, 
\begin{eqnarray}\label{collprobcal}
 p_{\rm{co}}=&& \sum_{x,z} P^2(X=x|Z=z)P(Z=z)\nonumber\\
            =&&\sum_{i=1}^4P^2(X=0|Z=P_i)P(P_i)+\nonumber\\
                   && P^2(X=1|Z=P_i)P(P_i)\nonumber\\
            =&&1/2[~\sum_{i=1}^4P^2(X=0_{1}|Z=P_i)P(P_i)+\nonumber\\
                 && P^2(X=1_{1}|Z=P_i)P(P_i)~]\nonumber\\  
                &&+1/2[~\sum_{i=1}^4P^2(X=0_{2}|Z=P_i)P(P_i)+\nonumber\\
                && P^2(X=1_{2}|Z=P_i)P(P_i) ].
    \end{eqnarray}

 The final step is to account for the fact that the bit detected can correspond to either the first phase difference, or the second one. The probabilities required to calculate Eq.~{\eqref{collprobcal}} are given in Table \ref{Table2}, and the average collision probability is found to be $0.72$.
Now, let the probability of correct detection by Eve be denoted as $p_{\text{corr}}$, and the fraction of bits intercepted by Eve as $n_{\text{int}}^{\text{Eve}}$. To maintain the error rate at Bob's side, the fraction of bits Eve intercepts is,
\begin{equation}
    (1-p_{\text{corr}}) n_{\text{int}}^{\text{Eve}} = e_b.
\end{equation}
We found from the SDP that $p_{\text{corr}}=0.75$ for the minimum error discrimination of the $3$-pulse DPS states. This indicates that the fraction of bits that can be intercepted by Eve is $4e_b$. Out of these $4e_b$ pulses intercepted only two-thirds of $4e_{b}$ are detected by Bob in either the second or in the third time bin. Thus for a fraction of $\left((2/3)\cdot 4e_b\right)$ pulses the collision probability is $0.72$ while for the remaining ($1-\left(2/3\right)4e_b$) fraction of pulses, the collision probability is half since they are unknown to Eve. The key rate in the presence of MED is thus given as,
\begin{multline}
R_{\text{MED}} = \frac{2}{3} p_{\text{click}} \left[ -\left(\frac{2}{3}\cdot 4e_b\right)\log_{2} 0.72 
+ \left(1-\frac{2}{3}\cdot4e_b\right)-f(e_b)h(e_b) \right]\label{med_rate}
\end{multline}

 The key rate is calculated for the following implementation parameters: Channel loss of $0.2$ dB/km, dark count probability of $10^{-6}$, detector efficiency of $10\%$, and $f(e_b) = 1.16$.

\begin{figure}[t]
    \centering
    \includegraphics[width=0.75\columnwidth]{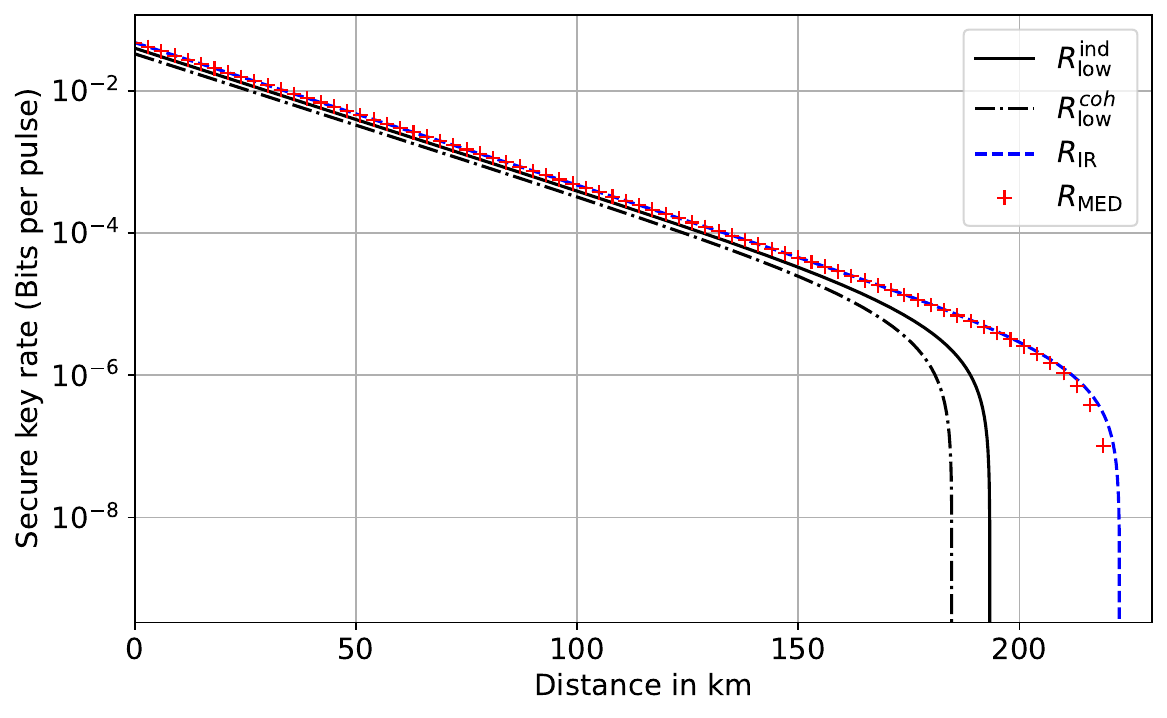}
    \caption{Secure key rate vs distance plot for the single-photon $3$-pulse DPS protocol in the presence of intercept and resend attack $(R_{\rm IR})$, minimum error discrimination attack $(R_{\rm MED})$, $(R_{\rm low}^{\rm ind})$, the theoretical lower bound on the secure key rate against individual attack and  $(R_{\rm low}^{\rm coh})$, the theoretical lower bound on the secure key rate against coherent attack.}
    \label{fig:skr}
\end{figure}

\begin{figure}[t]
   \centering
    \includegraphics[width=0.75\columnwidth]{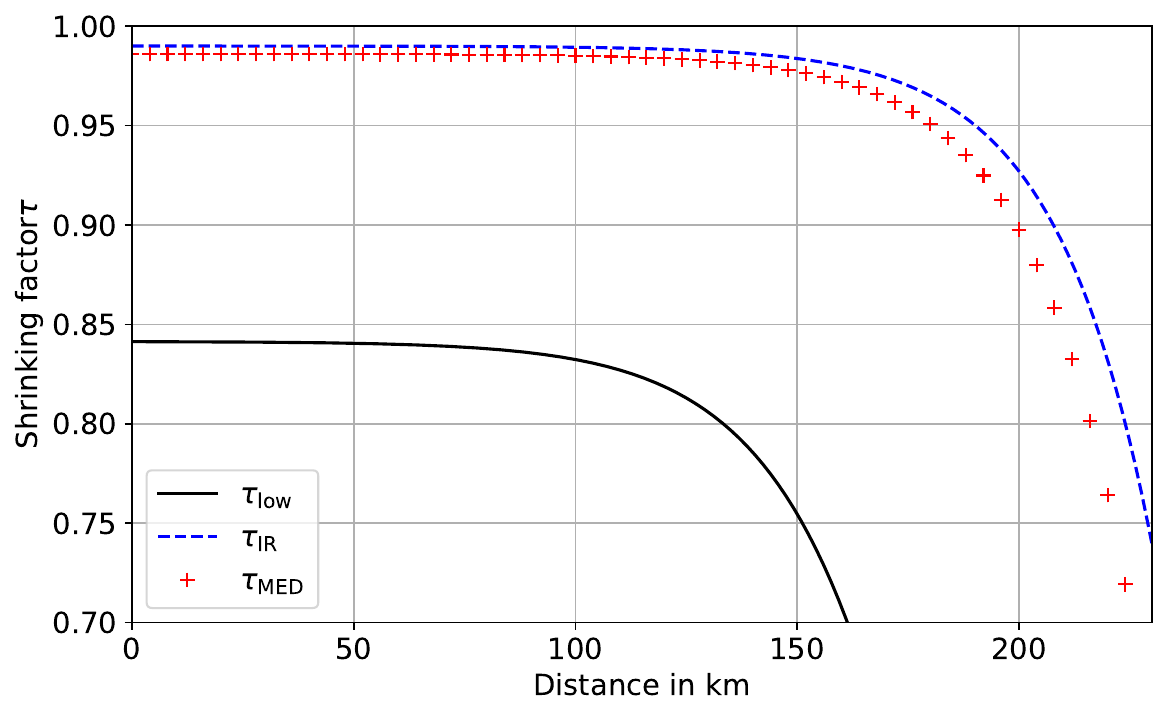}
    \caption{Shrinking factor at different channel lengths corresponding to the lower bound $(\tau_{\rm low})$, IR attack $(\tau_{\rm IR})$ and MED attack $(\tau_{\rm MED})$. A channel loss of $0.2$dB/Km and a detector efficiency of $10$\% are used to find the bit error rate and subsequently the shrinking factor.}
    \label{fig:IR_MED_Shrinking}
\end{figure}

We plot the calculated secure key rate for the single-photon 3-pulse DPS protocol in Fig.\ref{fig:skr}. In particular, the key rates in the presence of IR attack\cite{InterceptResend2017}, MED attack (Eq.\ref{med_rate}), as well as the two theoretical lower bounds in Eqs.(\ref{low_ind}) and (\ref{low_coh}) are plotted.
 Since the key rates are of the same order for different attacks, key rate plots which use a semi-log scale do not bring out the difference (Fig.\ref{fig:skr} ) between the lower bounds and the actual attacks. Therefore we also plot the shrinking factor to characterize the relative effectiveness of the various attacks.  
We plot in Fig. \ref{fig:IR_MED_Shrinking} the shrinking factor,
\begin{equation}
   \tau_{\rm~ MED} =  -\left(\frac{2}{3}\cdot 4e_b\right)\log_{2} 0.72 + \left(1-\frac{2}{3}\cdot4e_b\right),\label{tau_med}
\end{equation}
and compare it with the shrinking factor corresponding to the lower bound and IR attack. We note that the MED attack does only marginally better than the IR attack and is far from the theoretical bounds. This suggests that we can tolerate more error and get a higher key rate if we are to restrict Eve to MED attacks. In terms of the critical QBER, IR gives a value of $23\%$ while for MED the critical QBER is $19.9\% \approx 20\%$.

In the $4$-pulse case, Eve has to discriminate between $8$ different states and send the correct one to Bob. A POVM of $8$ elements has to be set up, and the probability of the right identification maximized. This SDP problem gives out that each state is correctly identified with a probability of $0.5$. Thus, in the $4$-pulse case, the minimum error attack is not a strong attack. Similarly, in the $5$-pulse case, the number of states goes to $16$ and the probability of right identification drops further down. Hence we conclude that an MED attack using this formulation is not powerful in the case of $n>3$ pulses. We also note that the structure of the POVM elements in the $4$-pulse and the $5$-pulse case looks similar to the $3$-pulse case because of the symmetry in the DPS states.   

Going beyond the ideal, single-photon protocol, we may ask a similar question about Eve's optimal discrimination strategy for $3$-pulse DPS protocol using weak coherent states (WCS). The states sent by Alice in a WCS-based DPS protocol are of the form,
$$|\psi\rangle_{\pm,\pm} = |\alpha\rangle_{k} \otimes |\pm\alpha\rangle_{k+1} \otimes |\pm\alpha\rangle_{k+2} ,$$ 
where $|\alpha|^{2}$ is the mean photon number of the WCS source. These states being linearly independent allows for an unambiguous state discrimination (USD) attack. The secure key rate for a WCS $3$-pulse DPS protocol subject to a USD attack is discussed in  Sec.~\ref{sec:WCS}.


\section{Cloning attack}\label{sec:cloningattack}
We now allow the eavesdropper to have a Quantum Cloning Machine (QCM), and she tries to clone the single-photon state sent to Bob. Since the theory of quantum mechanics does not allow for perfect cloning \cite{Wootters1982}, the cloned state is not a perfect copy and the initial state is disturbed in some manner. In this section, we find the bit error rate due to such imperfect cloning.

\subsection{Cloning as an SDP problem}

First we consider a cloning attack described by a quantum channel $\Phi$, which takes a state $\rho = |\psi\rangle\langle \psi|  \in \mathrm{D}(\mathcal{X})$ to the state $\Phi(\rho) \in \mathrm{D}(\mathcal{Y} \otimes \mathcal{Z})$. Here, $\mathrm{D}$ refers to the set of density matrices on the corresponding Hilbert spaces; $\mathcal{X}$, $\mathcal{Y}$, $\mathcal{Z}$ correspond to the sender (Alice) Hilbert space and the two receivers (Bob and Eve) respectively. The channel $\Phi$ must correspond to a completely positive and trace-preserving linear mapping of the form $\Phi: \mathrm{D}(\mathcal{X}) \rightarrow \mathrm{D}(\mathcal{Y} \otimes \mathcal{Z}) .$ Suppose the state sent is indexed $i$, a good measure that can quantify the extent of cloning is given by the fidelity, $\left\langle\psi_{i} \otimes \psi_{i}\left|\Phi\left(\left|\psi_{i}\right\rangle\left\langle\psi_{i}\right|\right)\right| \psi_{i} \otimes \psi_{i}\right\rangle $\cite{molina2012optimal}. 
Averaging over the possible choices of $i$, we get the average fidelity function corresponding to the cloning map $\Phi$ as,
\begin{equation}
\sum_{i=1}^{N} p(i)\left\langle\psi_{i} \otimes \psi_{i}\left|\Phi\left(\left|\psi_{i}\right\rangle\left\langle\psi_{i}\right|\right)\right| \psi_{i} \otimes \psi_{i}\right\rangle. \label{eq:avg_fid}
\end{equation}
The optimal cloning strategy can then be obtained by maximizing this fidelity function over all valid channels $\Phi: \mathrm{D}(\mathcal{X}) \rightarrow \mathrm{D}(\mathcal{Y} \otimes \mathcal{Z})$. This optimization problem can again be represented by a semidefinite program~\cite{WatrousBook, molina2012optimal}.

The formulation makes use of the Choi-Jamiołkowski representation $J(\Phi) \equiv X$ of a given channel $\Phi$. The SDP is given as \cite{molina2012optimal}

\begin{align}
    \text{maximize} \qquad & \langle Q, X \rangle \nonumber \\ 
  \text{ subject to} \qquad &
   \text{Tr}_{\mathcal{Y}\otimes\mathcal{Z}}(X) = I_{\mathcal{X}},\\ 
 & \mathcal{X} \in \operatorname{Pos}(\mathcal{Y} \otimes \mathcal{X} \otimes \mathcal{Z}), \nonumber
\end{align}\label{cloningSDP}
%
where
$$
Q=\sum_{i=1}^{N} p(i)\left|\psi_{i} \otimes \psi_{i} \otimes \overline{\psi_{i}}\right\rangle\left\langle\psi_{i} \otimes \psi_{i} \otimes \overline{\psi_{i}}\right|.
$$
 {Here, $\overline{\psi}$ refers to taking complex conjugation with respect to the standard basis.}

\subsection{Optimal Cloning Attack and Bit Error Rate for $3$-pulse DPS states} \label{sec:optimalcloning}
We solve the SDP formulated in Eq. \eqref{cloningSDP} for the specific case of the $3$-pulse DPS states given in Eq. \eqref{eq:eq1}. The optimal value for the average fidelity function defined in Eq.~\eqref{eq:avg_fid} is evaluated to be $0.78$.
The fidelity between the original state sent by Alice and the two cloned states is $0.81$ for all the states $|\psi\rangle_{\pm,\pm}$ in Alice's ensemble. Using the cvx solver, we also obtain the map $\Phi$ acting on Alice's states, from the Choi matrix.  The details of the Choi matrix obtained and the details of the corresponding map are discussed in Appendix~\ref{sec:Appendix A}. Interestingly, the map on Alice's states corresponding to the optimal cloner turns out to be a qutrit depolarizing map. 

The bit error rate at Bob's side due to the application of the optimal cloning machine can be found using the probability of detecting in the constructive  (destructive) port when the destructive (constructive) port is supposed to click. Here, we estimate the bit error rate for the optimal cloning attack as follows. We first note that the state received by Bob after the cloning attack is a mixed state. We write down the density matrix corresponding to this mixed state as a convex combination in an appropriate basis and calculate the probability of a wrong detection.\\

\begin{figure}
    \centering
    \includegraphics[width=0.75\columnwidth]{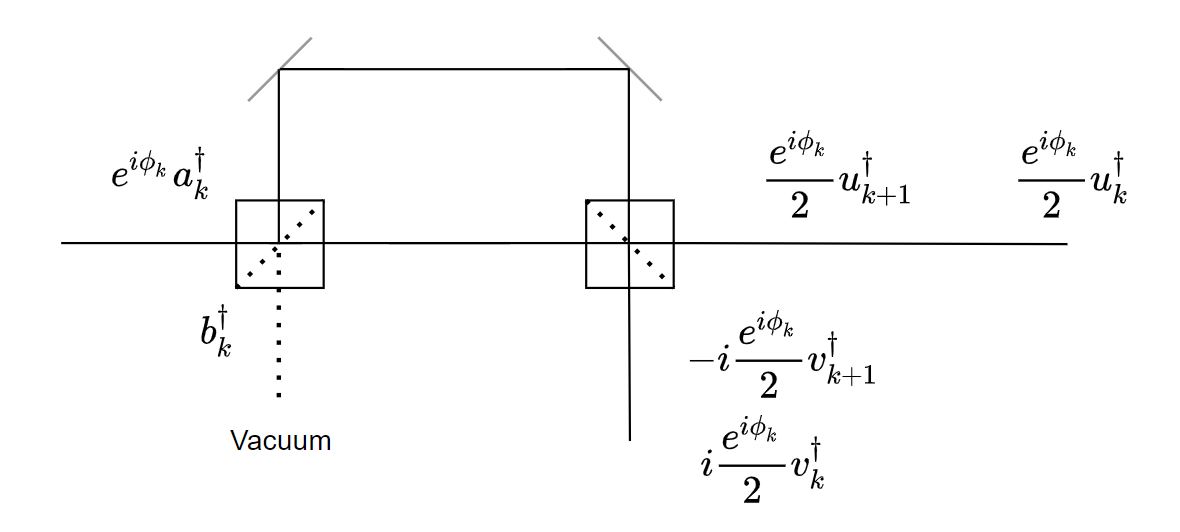}
    \caption{Input-output relation of Bob's asymmetric MZI. $a^\dagger_k$, $b^\dagger_k$ represents the creation operator at the input ports at $k_{\rm}$ time and $u^\dagger_k$,$v^\dagger_k$ the creation operators at the output port.}
    \label{MZI}
\end{figure}

Consider the density matrix $\rho_{+,+}$ corresponding to Alice's input state $|\psi\rangle_{+,+}$, of the form,
\begin{equation}\rho_{+,+}= \frac{1}{3}\left(\begin{array}{ccc} 1 & 1 & 1\\ 1 & 1 & 1\\ 1 & 1 & 1 \end{array}\right)\label{rho1}.\end{equation}
After the action of the QCM, the reduced density matrix available to Bob and Eve is \begin{equation}\tilde{\rho}_{+,+} = \left(\begin{array}{ccc} \dfrac{1}{3} & 0.23 & 0.23\\ 0.23 & \dfrac{1}{3} & 0.23\\ 0.23 & 0.23 & \dfrac{1}{3} \end{array}\right) .\label{bobket1}\end{equation}
An error occurs in Bob's detector if there is a detection in the wrong port. 
Recall that the detector at Bob's side is an asymmetric MZI with the input-output relation shown in Fig. \ref{MZI}.
The original state $|\psi\rangle_{+,+}$ sent by Alice in the $k^{\rm th}$ slot can be written in terms of the photon creation operators, as,
\begin{equation}
    |\psi\rangle_{+,+} = \frac{1}{\sqrt{3}}\left[a_{\rm k+2}^\dagger + a_{\rm k+1}^\dagger + a_{\rm k}^\dagger\right]|0\rangle .
\end{equation} 
Here, as before, $a_k^\dagger$ denotes the photon creation operator at the $k$th time slot.
After $|\psi\rangle_{+,+}$ goes through the MZI, the output state is, 
\begin{equation}
    \frac{1}{\sqrt{3}}\left[\frac{u_{\rm k}^\dagger}{2} + \frac{iv_{\rm k}^\dagger}{2}+ u_{\rm k+1}^\dagger + u_{\rm k+2}^\dagger + \frac{u_{\rm k+3}^\dagger}{2} - \frac{iv_{\rm k+3}^\dagger}{2}\right]|0\rangle.
\end{equation}
Thus, in the ideal case, only the constructive port clicks at the $(k+1)$ and $(k+2)^{\rm th}$ time slots.

Now let us look at what happens when the cloned state $\tilde{\rho}_{+,+}$ is received by Bob's asymmetric MZI. To calculate the probabilities of detection at different ports and time instances, we  
diagonalise $\tilde{\rho}_{+,+}$ and apply the MZI transformation on its eigenstates. 
The orthonormal basis that diagonalizes the cloned state $\tilde{\rho}_{+,+}$ is 
\begin{eqnarray}
    |e_{1}\rangle &=&  \frac{1}{\sqrt{3}}\left[|1\rangle + |2\rangle + |3\rangle\right] \nonumber \\ |e_{2}\rangle &=& \frac{1}{\sqrt{6}}\left[|1\rangle + |2\rangle -2 |3\rangle\right] \nonumber \\ |e_{3}\rangle &=& \frac{1}{\sqrt{2}}\left[|1\rangle - |2\rangle\right] .
\end{eqnarray} 

\noindent The spectral decomposition of $\tilde{\rho}_{+,+}$ can be written as,
\begin{equation}
 \tilde{\rho}_{+,+} = 0.81|e_{1}\rangle\langle e_{1}| + 0.095 |e_{2}\rangle\langle e_{2}| + 0.095 |e_{3}\rangle\langle e_{3}|. \label{eq:spec}
\end{equation}
The second and third terms in Eq.~\eqref{eq:spec} cause a detection in the destructive port. This erroneous click is observed due to $|e_{2}\rangle$ at the $(k+2)^{\rm th}$ slot and $|e_{3}\rangle$ in both slots $(k+1)$ and $(k+2)$. After passing the cloned state through the MZI at Bob's side, the probability of error (in this case - the probability of clicking in the destructive port in the second and third-time bin) can be calculated as,
\begin{equation}
   \text{BER} = 0.095 \left(\frac{3}{8}\right) + 0.095 = 0.13.
\end{equation}

Repeating the analysis for all four DPS states, we find that the probability of wrong detection corresponding to each of the cloned states is $0.13$. Once again, like in the case of the MED attack, an eavesdropper will try to clone only a number $n_{\text{int}}^{\text{Eve}}$ of states such that the error introduced due to eavesdropping matches the error introduced due to the channel.
$$p_{\text{error}}\cdot n_{\text{int}}^{\text{Eve}} = e_b.$$
If the QBER  after the sifting at Bob's side is expected to be $e_b$, Eve can intercept a fraction of $7.69e_b$ bits and make sure she does not cross the QBER threshold $e_b$.

\begin{figure}
    \centering
    \includegraphics[width=0.75\columnwidth]{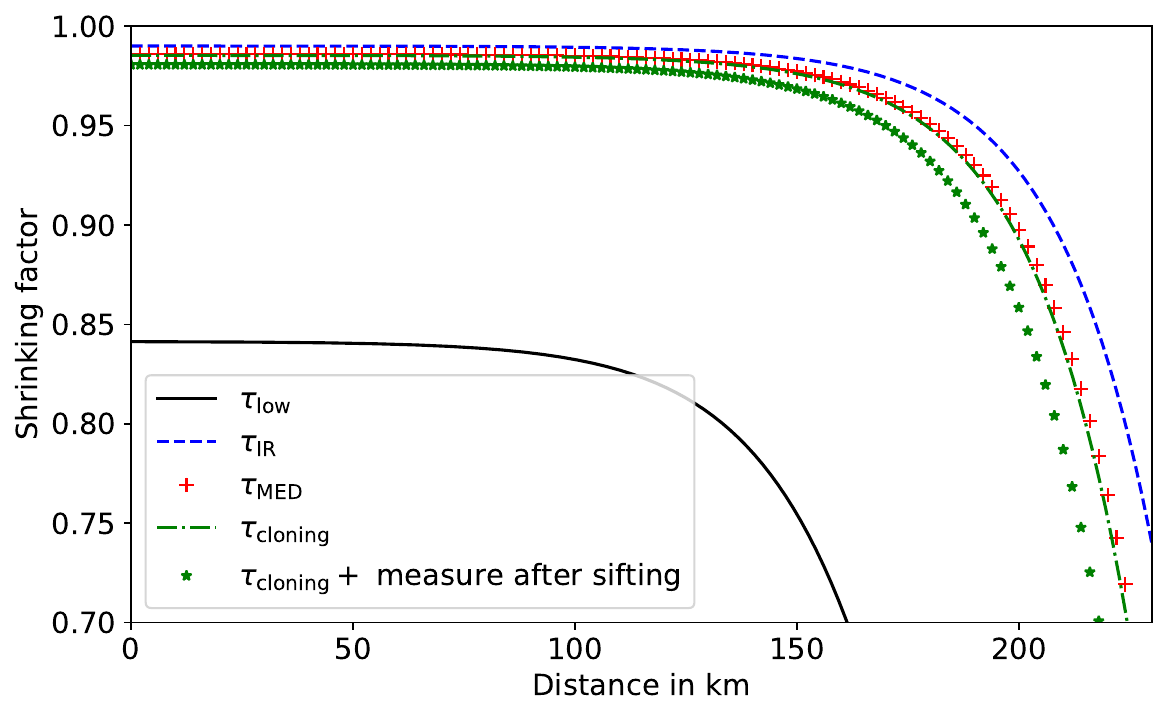}
    \caption{Shrinking factor corresponding to the lower bound $(\tau_{\rm low})$, IR attack $(\tau_{\rm IR})$, MED attack $(\tau_{\rm MED})$ and cloning attack $(\tau_{\rm cloning})$ at different channel lengths. A channel loss of 0.2dB/Km and a detector efficiency of 10\% is used to find the bit error rate.}
    \label{fig:IR_MED_cloning_shrink}
\end{figure}

\subsection{Collision probability and key rate with cloning attack}

  To quantify the maximum information available to an adversary, we let Eve do a minimum error discrimination of the states reaching her after the cloning machine. We note here that since the cloning map acts like a depolarizing channel on Alice's state, the optimized POVMs are the same as those found for the pure DPS states. Since the states reaching Eve and Bob are mixed, the MED success probability is relatively less than in the case where Eve does a MED directly on the states sent by Alice. Nevertheless, this strategy does marginally better since the error introduced at Bob due to cloning is less than the direct MED, thus allowing for more bits to be attacked. Denoting the POVM elements for MED of the cloned state as CP$_i$, the SDP results for the right identification are as given in Table~\ref{clonedMED}. The collision probability is found for these POVM as in the case of the MED attack and is found to be 0.61. The corresponding secure key rate is given as  

\begin{align}
R_{\text{cloning}} & = \frac{2}{3} p_{\text{click}} \left[ -\left(\frac{2}{3} \cdot 7.69e_b\right)\log_{2} 0.61 
+  \left(1-\frac{2}{3} \cdot 7.69e_b\right) -f(e_b)\cdot h(e_b) \right] , \\
\tau_{\text{cloning}}  & =  \left[ -\left(\frac{2}{3}\cdot7.69e_b\right)\log_{2} 0.61  
+  \left(1-\frac{2}{3} \cdot 7.69e_b\right) \right] .
\end{align}

\begin{table}[ht]
\caption{Measurement probabilities for the $3$-pulse DPS states after the application of the optimal cloning map. $\{CP_i\}$ represents   the POVM elements}
\label{clonedMED}

\begin{tabular}{  c  c  c  c  c  }
\toprule
 & $\rm{CP_1}$& $\rm{CP_2}$&$\rm{CP_3}$ &$\rm{CP_4} $ \\
\midrule
 $\psi_{+,+}$&$0.607$& $0.131$ & 0.131 & 0.131 \\

 $\psi_{+,-}$&0.131& $0.607$ & 0.131& 0.131 \\
 
 $\psi_{-,+}$&0.131&0.131&0.607&0.131\\
 
 $\psi_{-,-}$&0.131 & 0.131 & 0.131& 0.607  \\
 \botrule
\end{tabular}

\end{table}
We plot in Fig.~\ref{fig:IR_MED_cloning_shrink}, the shrinking factor for the different attacks discussed so far. The cloning attack is seen to be only as powerful as the IR and MED attacks. The critical QBER in the presence of a cloning attack is $19.5\% \approx 20 \%$. Our detailed analysis of the different individual attacks thus indicates that the critical QBER of $6\%$ imposed by the lower bound is much more restrictive on the tolerable QBER.

We finally note that in the presence of a \emph{perfect} (noiseless) quantum memory, the eavesdropper can perform a modified cloning attack where she waits until Bob sifts the key and reveals which of the two phases he detected. With this sifting  information, Eve can measure only the sifted phase from her copy of the cloned state. This would indeed be a stronger attack compared to performing MED immediately after the cloning operation. The shrinking factor for this modified cloning attack is also plotted in Fig.~\ref{fig:IR_MED_cloning_shrink}.


\section{A QCM based on unitary transformation}\label{sec:unitarycloning}
In this section, we look at a simpler quantum cloning machine based on a unitary transformation~\cite{Buzek98} on a tripartite system $\mathcal{H}_A \otimes \mathcal{H}_B \otimes \mathcal{H}_C$, where $\mathcal{H}_A$ denotes the Hilbert space of the state to be cloned, $\mathcal{H}_B$ denotes the ancillary system into which the copy is made and $\mathcal{H}_C$ denotes the Hilbert space associated with the cloner. Suppose $\mathcal{H}_A$ and $\mathcal{H}_B$ are Hilbert spaces of dimension $d$. Let $\mathcal{B} \equiv\left\{|e_i\rangle_A, 1 \leqslant i \leqslant d\right\}$ denote a fixed orthonormal basis in $\mathcal{H}_A$. The symmetric QCM is then defined via the linear transformation \cite{Buzek98, Mitra_2021},
\begin{equation}
|e_i\rangle_A|0\rangle_B|X\rangle_C  \rightarrow p|e_i\rangle_A|e_i\rangle_B\left|X_i\right\rangle_C 
 +q \sum_{j \neq i}~\left (|e_i\rangle_A|e_j\rangle_B+|e_j\rangle_A|e_i\rangle_B\right )~\left|X_j\right\rangle_C . \label{cloningUnitary}
\end{equation}
where, $|X\rangle_C$ is a fixed (normalized) state of system $\mathcal{H}_C,|0\rangle_B$ is a blank ancilla state which is transformed to a clone of $|e_i\rangle_A$, and $\left\{\left|X_i\right\rangle_C\right\}$ are the fixed orthonormal basis vectors of cloning machine. The state is symmetric in the first two systems. The real coefficients $p$ and $q$ must satisfy the following relation to ensure unitarity of the QCM transformation:
\begin{equation}
p^2+2(d-1) q^2=1 .
\end{equation}
We use the analytical solution given in Appendix B of \cite{Mitra_2021} to find the best cloning fidelity for the $3$-pulse DPS states. 
 Based on the analytical solution given in \cite{Mitra_2021}, we find that the optimal $q$ in eq. (\ref{cloningUnitary}) is $0.23$. The average fidelity of the cloned copies is found to be $0.78$. 
 
\begin{figure}
    \centering
    \includegraphics[width=0.75\columnwidth]{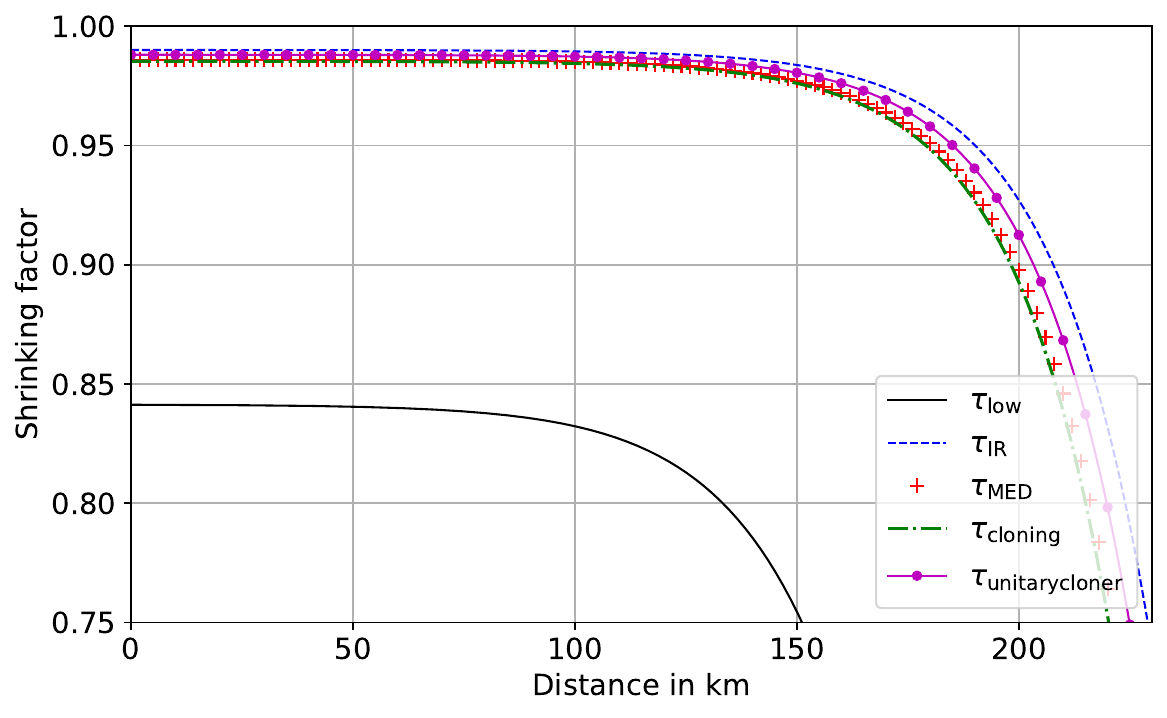}
    \caption{Shrinking factor corresponding to the lower bound $(\tau_{\rm low})$, IR attack $(\tau_{\rm IR})$, MED attack $(\tau_{\rm MED})$, optimal cloning $(\tau_{\rm cloning})$ and unitary cloning $(\tau_{\rm unitarycloner})$ attacks at different channel lengths. A channel loss of $0.2$dB/Km and a detector efficiency of $10$\% is used to find the bit error rate.}
    \label{fig:IR_MED_cloning_neclon_Shrinking}
\end{figure}

 We now use the same approach as in the case of the optimal cloning attack discussed in Sec.~\ref{sec:cloningattack} to estimate the bit error rate due to the unitary cloning attack. As discussed in Appendix~\ref{unitarycloner}, the BER in the case of the unitary cloner evaluates to $ 0.15$. Thus the fraction of bits that can be intercepted by the eavesdropper to perform the unitary cloning attack is $6.66~ e_b$. 
After cloning, Eve is allowed to do a minimum error discrimination of the states just as in the case of the optimal cloning attack of the previous section. On running the SDP, it is found that the success probability of correct discrimination is $0.603$. Thus the key rate in the presence of this attack is \begin{multline}
     R_{\rm UCloning} = -\frac{2}{3}  p_{\rm click}\left[-\frac{2}{3}6.66e_b\cdot \log_{2} 0.603 + \left(1-\left(\frac{2}{3}\right)6.66e_b\right)+f(e_b)\cdot h(e_b)\right].
\end{multline}
Fig.~\ref{fig:IR_MED_cloning_neclon_Shrinking} shows the shrinking factor of the DPS protocol under all the attacks discussed so far. We see that cloning based on unitary transformation is not optimal compared to map-based cloning. The fidelity of cloning is lower and hence the ability to distinguish the cloned state by Eve is low. Moreover, none of the sophisticated attacks are close to the lower bound.

\section{Finite size effects}
\begin{figure}[t]
    \centering
    \includegraphics[width=0.75\columnwidth]{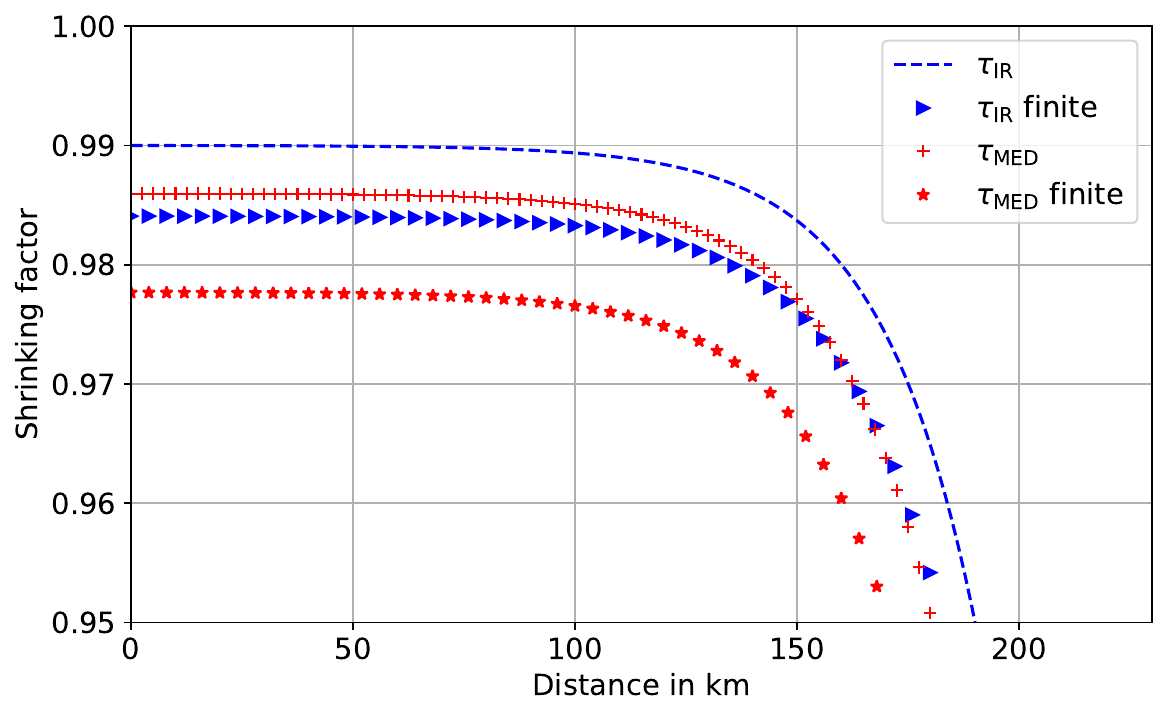}
    \caption{Shrinking factor corresponding to  IR $(\tau_{\rm IR})$ and MED  $(\tau_{\rm MED})$attacks at different channel lengths. The effect of finite size can be seen in the key rate reduction. A similar effect is seen for the other attacks. A channel loss of 0.2dB/km and a detector efficiency of 10\% is used in Eq. \ref{eq:eb} to find the bit error rate.}
    \label{fig:finite}
\end{figure}
In many cases, finite-size effects have been widely studied for QKD protocols and are said to be the limiting factor. Here we include the finite size effects to the secure key rate against individual attacks. The three parameters that need to be considered are 
\begin{enumerate}
    \item The statistical fluctuations in parameter estimation (PE) (fluctuations in QBER), 
    \item Failure probability of error correction procedure,  
    \item Failure probability of privacy amplification (PA) protocol used. 
\end{enumerate}
The first one is dominant and is orders larger than the other two. Since the failure probability of the EC and PA procedures scale as the collision probability of the hash functions used. Therefore they are usually of the order of $10^{-12}$. To account for the finite size correction in the qubit error rate statistics, we use the tail bound derived in \cite{Korzh2015}. If $e_{\rm obs}$ denotes the observed QBER and $e_{\rm key}$ is the actual QBER of the key sifted, the bound is given as \cite{Korzh2015},
\begin{equation}
    e_{\rm key} \leq e_{\rm obs} + t ,
\end{equation}
where

\begin{multline}
t\left(n, k,e, \epsilon^{\prime}\right)  :=\sqrt{\frac{2(n+k) e(1-e)}{k n} \log \frac{\sqrt{n+k} C(n, k, e)}{\sqrt{2 \pi n k e(1-e)} \epsilon^{\prime}}}, \\
 C(n, k, e) :=\exp \left(\frac{1}{8(n+k)}+\frac{1}{12 k}-\frac{1}{12 k e+1}-\frac{1}{12 k(1-e)+1}\right)\nonumber .
\end{multline}
Here $n+k$ is the total number of bits sifted and $k$ is the bits used for parameter estimation.

The deviation $t$ is negligible for higher block lengths and we see visible effects for smaller ones. We plot in Fig.\ref{fig:finite} the effects for a block length of $n=10^{6}$, a sample of $k=10^{4}$ for parameter estimation and for confidence parameter $\epsilon^{\prime} = 10^{-9}$.

\section{Three-pulse DPS with a weak coherent source}\label{sec:WCS}
In this section we briefly introduce attacks on weak coherent source (WCS) based DPS. The 3-pulse DPS protocol with a weak coherent source, in place of single photon sources, is not a superposition state but rather a product state and is given as:
$$|\psi\rangle_{\pm,\pm} = |\alpha\rangle_{k} \otimes |\pm\alpha\rangle_{k+1} \otimes |\pm\alpha\rangle_{k+2} .$$

This protocol then looks like the pulse train DPS but with two bits of data encoded in the phase between three pulses. The USD attack is then an individual attack on the 3-pulse sequences carrying two bits of data. We assume that the eavesdropper has a local oscillator phase locked to Alice's source. Thus Eve can intercept each pulse and do unambiguous state discrimination to identify whether each pulse is $|\alpha\rangle$ or $|-\alpha\rangle$. The probability of successful discrimination is then given by the  Ivanovic-Dieks-Peres (IDP) limit \cite{DIEKS1988303},
\begin{equation}
    P_{\text{suc}} = 1 - |\langle\alpha |-\alpha\rangle| = 1-e^{-2\mu_\alpha}.
\end{equation}
where $\mu_\alpha = |\alpha^2|$.
This puts a tighter upper bound on the secret key rate. At a higher mean photon number, the USD attack has more probability for right identification. Since the protocol with coherent state becomes equivalent to a pulse train DPS, the strength of the USD attack is equivalent to the one in \cite{Curty_DPSUSDSequential} with a block size (M in \cite{Curty_DPSUSDSequential}) of three. With the block size three, the sequential attack \cite{Curty_DPSUSDSequential} in the 3-pulse DPS becomes an individual attack. Practical implementations of USD are now becoming more possible with currently available tools \cite{usdPracticalLupo}.    

 {Security of weak coherent pulse-train DPS has been studied against general individual attacks in Ref.~\cite{IndividualAttackSecurity, MGIA2024}, and against collective and sequential attacks in \cite{CombatingSeq2011, Kronberg_2014, Curty_DPSUSDSequential}. Sequential USD attacks can be avoided by slowly modulating the phase, as shown in Ref.~\cite{CombatingSeq2011}, since this prevents an eavesdropper from performing USD. In the next section, we examine how complete phase randomization, works as a countermeasure against USD attacks in 3 pulse DPS. }


\subsection{Phase randomized three-pulse DPS with weak coherent source}


\begin{figure}[h]
    \centering
    \includegraphics[width=1\linewidth]{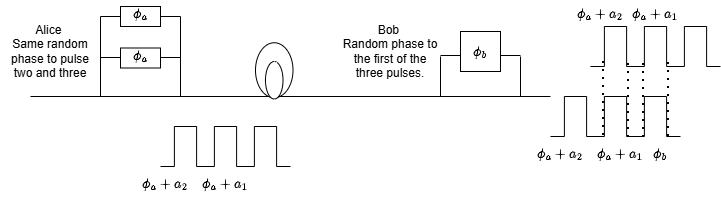}
    \caption{Alice applies a random phase $\phi_a$ in both her delay lines (same random phase for both delay lines, alternatively she can apply a random phase to pulse two and three after the delay line interferometer). This leads to the three pulses having phases as shown in the diagram. Bob applies a random phase to the first of the three pulses.}
    \label{PhaserandomizedDPS}
\end{figure}

In this section, we look at the effectiveness of introducing a random phase at both Alice and Bob's side as a measure to protect against cloning and USD attack. When the phase is randomized, one cannot carry out the above-mentioned attacks. Since the set of states to be discriminated or cloned has increased from just \{$|\alpha\rangle,|-\alpha\rangle $\} to states with random phases. The random phases are applied as follows:

\begin{enumerate}
   \item Alice introduces the same random phase $\phi_a$ in both her delay lines.
    \item Bob introduces a random phase $\phi_b$ in the delay line of the MZI as shown in Fig. \ref{PhaserandomizedDPS} in the first pulse of the three pulses sent by Alice.
\end{enumerate}

\noindent When the phases introduced by Alice and Bob do not match, it increases the probability of wrong detection and thus to QBER. Following \cite{Lucamarini2018}, we look at the error introduced when both Alice and Bob choose a random phase that is $\delta$ angle apart. 

\begin{figure}[b]
  \centering
    \includegraphics[width=0.65\columnwidth]{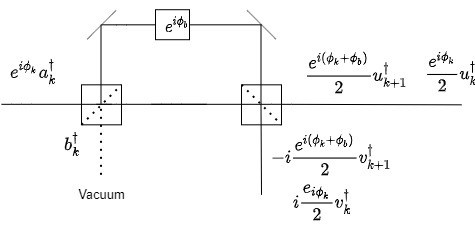}
  \caption{Input-output relation for asymmetric MZI with phase.}
    \label{MZI_withphase}
\end{figure}

\noindent Consider the case when Alice sends the state \{$|e^{i\phi_a}\alpha\rangle_{k+2}\otimes |e^{i\phi_a}\alpha\rangle_{k+1} \otimes |\alpha\rangle_{k} $\}. In the absence of any random phase, the output should ideally be a click in the constructive port at time instants $2$ and $3$. In the presence of a random phase by Alice and Bob, it can be calculated using the input-output relation shown in Fig.\ref{MZI_withphase}, that the state reaching the destructive port for the second time instant is
$$|ie^{i\phi_a}(1-e^{i\delta})\rangle,$$ where $\delta = \phi_b - \phi_a$. The probability of detection in the destructive port is the QBER and is given as 

\begin{equation}
  \text{QBER} =  1- e^{-\alpha^2\sin^2 \delta/2} \approx \alpha^2\sin^2 \delta/2. 
\end{equation}

Thus, the QBER introduced due to phase randomization depends on the mean photon number $(\alpha)$ and the phase difference $(\delta)$ between the random phases. The higher the phase difference, the higher the QBER. Hence we look at a post-selection method in which Alice and Bob announce and choose only phases that lie within a small range. Say the interval $[0,2\pi]$ is cut into $M$ slices. The pulses in which both Alice and Bob select random phases within the same slice is only used for the secret key. We find the average QBER when $\delta$ lies within a slice and find that the QBER approaches zero as $M$ is greater than 15. We use 16 slices and hence the maximum variation between the random phase or $\delta$ is $\pi/8$. We now use this to calculate the QBER and plot the secret key rate for a mean photon number of 0.4.
\begin{figure}[]
\centering
   \includegraphics[width=0.75\columnwidth]{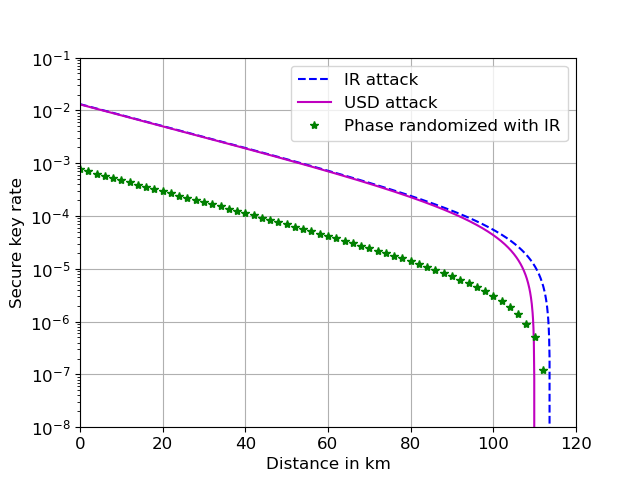}
    \caption{Secure key rate vs distance for weak coherent state-based DPS protocol in the presence of IR attack, USD attack, and for protocol with phase randomization in the presence of IR attack.}
    \label{PlotFinal}
\end{figure}
From Fig \ref{PlotFinal}, it can be noted that the curve with phase randomization is lower than the curve with USD attack since there is a further sifting factor of $1/16$ multiplied for Alice and Bob to choose the same phase slice. If we ignore the sifting factor, phase randomization provides a marginal increase in distance at high mean photon numbers.

\section{Conclusion}\label{sec:conclusion}

This study provides a practical assessment of the security of DPS QKD protocols against explicit, physically implementable individual attacks. Our results show that the analyzed strategies such as minimum error discrimination and quantum cloning remain far from theoretically optimal, and serve to establish meaningful benchmarks that reflect realistic adversarial capabilities. The findings also highlight the gap between such explicit attacks and information-theoretic optimality, as summarized in Table~\ref{BER}.

Experimental methods that can implement coherent state discrimination attacks have been recently reported in the literature~\cite{cui2024superadditivecommunicationgreenmachine, usdPracticalLupo}. Identifying experimental procedures that can similarly be used to implement the optimized POVM measurement and cloning map constructed here would be a particularly intriguing avenue for further research.

{We acknowledge that a full experimental validation of the proposed optimised attacks is beyond the scope of the present manuscript. To provide an appropriate comparative assessment instead of experimental data, we have incorporated detailed benchmarking against established results in the open literature, including standard individual attacks (beam-splitting and intercept–resend) and, for the weak coherent state implementation, the unambiguous state discrimination (USD) attack. These theoretical comparisons place our performance and security bounds within the widely accepted QKD framework and offer a meaningful validation of the approach. Nonetheless, experimental implementation and empirical benchmarking remain important limitations of the current work and constitute immediate directions for future studies.}

From a system engineering perspective, these results help define practical security margins and inform implementation choices, particularly in contexts where key lifetimes are limited and system noise is constrained by hardware. Future research should focus on extending these characterizations to incorporate real device imperfections and on experimentally demonstrating the feasibility of the optimized attacks studied. Such endeavors will be crucial for bridging the gap between theoretical security guarantees and the realities of deployed quantum communication systems.

\begin{table}[ht!]
\centering
    \caption{Protocols, attacks, and corresponding figure of merit. $\mu$ - mean photon number, $e$ - error rate due to implementation. The probability of right identification of a state when attacked - $P_\text{corr}$.}\label{BER}
\begin{tabular}{p{0.25\columnwidth}p{0.25\columnwidth}p{0.1\columnwidth}p{0.1\columnwidth}p{0.1\columnwidth}l}
\hline
\textbf{Protocol considered} & \textbf{Attack} & \textbf{QBER post- } & \textbf{$P_\text{corr}$} & \textbf{Critical QBER}  &Reference\\ 
& & \textbf{attack} & & \textbf{ (in \%)} &\\
\hline
\multirow{5}{*}{Single photon 3-pulse DPS}  & IR attack& 0.33                                   &                     & 23                              &\cite{InterceptResend2017}\\   
                                            & MED attack                                                                                   & 0.25                                   & 0.75                & 20                              &This work\\   
                                            & Optimal cloning attack                                                                       & 0.13                                   & 0.78                & 20                              &This work\\   
                                            & Individual attack bound &                                        &                     & 6                               &\cite{IndividualAttackSecurity}\\   
                                            & Coherent attack bound &                                        &                     & 4                               &\cite{UnconditionalSecurity_PRL}\\ \hline
\multirow{2}{*}{WCS 3-pulse DPS\footnotemark[1]}           & IR attack & 0.33                                   &                     &             12.5                    &\cite{InterceptResend2017}\\   
                                            & USD attack                                                                                   &                                        & $1-e^{-2\mu}$         &             10.5                    &\\  
                                            & Phase randomized
                                                                         &                       & &12
                                                 &This work\\ \hline
\multirow{3}{*}{Single photon  $n$-pulse DPS} & IR attack  & $0.5({\frac{n-1}{n}})^2$        &                     &                                 &\cite{InterceptResend2017}\\   
                                            & \multirow{2}{*}{\begin{tabular}[c]{@{}l@{}}MED attack (n=4)\\ MED attack (n=5)\end{tabular}}\footnotemark[2] &                                        & 0.5                 &                                 &This work\\   
                                            &                                                                                              &                                        & 0.31                &                                 &This work\\ \hline
\multirow{2}{*}{WCS Pulse train DPS }                        & IR attack                                                                                    & 0.25                                   &                     &                                 &\cite{DPSSecurity}\\  
                                            & USD attack  &                                        & $1-e^{-2\mu}$         &                                 &\cite{Curty_DPSUSDSequential}\\ \hline
\end{tabular}
\footnotetext[1]{We consider a channel length of 100~km and a mean photon number of 0.4 to calculate the critical QBER}
\footnotetext[2]{n-pulse MED: Critical QBER is not calculated since the attack is not implemented, the probability of correct detection is very low that the attack cannot be introduced for $n>3$ cases.}
\end{table}

\backmatter
\bmhead{Data Availability}

No new experimental or observational data were generated in this study. The key rate calculations and related data analyses were performed through theoretical modeling and numerical simulations. The code used for the optimization of the attacks and the key rate analysis is publicly available~\cite{valliamairm_2026_19617025}.

\bmhead{Code Availability}
 The code used for the optimization of the attacks and the key rate analysis is publicly available~\cite{valliamairm_2026_19617025}.

\bmhead{Acknowledgements}

This research was supported in part by MeitY (sanction 13(33)/2020-CC\&BT) and by a grant from the Mphasis F1 Foundation to the Centre for Quantum Information, Communication, and Computing (CQuICC). \\

\section*{Declarations}

{\noindent\textbf{Ethical approval:} \\
Not applicable.}

\noindent{\textbf{Consent to participate:} \\
Not applicable.}

\noindent{\textbf{Consent to publish:} \\
Not applicable.}

\noindent {\textbf{Conflict of interest:}\\
The authors declare no competing interests.}

\noindent {\textbf{Author contributions}\\
V.R. performed the main calculations of the results presented in the manuscript. A.P. and P.M. helped to formulate the problem statements studied here. All three contributed towards the writing of the main text and reviewed the manuscript.}

\noindent {\bf Funding:} This work was supported by MeitY (sanction 13(33)/2020-CC \& BT) and Mphasis F1 Foundation to CQuICC. \\

\noindent {\bf Corresponding author:} Prabha Mandayam\\
{\bf Email address:} prabhamd@physics.iitm.ac.in

\begin{appendices}

\section{Choi matrix and Optimal Cloning Map}\label{sec:Appendix A}

Here we describe the action of the optimal cloning attack described in Sec.~\ref{sec:optimalcloning}, on the $3$-pulse DPS states. To find the final cloned state shared between the receiver (Bob) and the eavesdropper (Eve), we calculate the trace, $\operatorname{Tr}(X\rho)$, where $X$ is the Choi matrix obtained by solving the SDP in Eq.~\eqref{eq:sdp_main} and $\rho$ is the density matrix corresponding to the state to be cloned. For example, evaluating $\operatorname{Tr}(X\rho_{+,+})$ gives us the two-qutrit state that is shared between Bob and Eve after the cloning map. Finally, to obtain the state reaching Bob, we simply trace out Eve's qutrit. On doing this we notice that the density matrix initially with Alice,
\begin{equation}\rho_{+,+}= \frac{1}{3}\left(\begin{array}{ccc} 1 & 1 & 1\\ 1 & 1 & 1\\ 1 & 1 & 1 \end{array}\right),\end{equation}
gets transformed to, \begin{equation}
\tilde{\rho}_{+,+} = \left(\begin{array}{ccc} \dfrac{1}{3} & 0.23 & 0.23\\ 0.23 & \dfrac{1}{3} & 0.23\\ 0.23 & 0.23 & \dfrac{1}{3} \end{array}\right).\end{equation}

Similarly, the action of the optimal cloning map on the other states can be evaluated as follows.
\begin{equation}\rho_{+,-}= \frac{1}{3}\left(\begin{array}{ccc} 1 & 1 & -1\\ 1 & 1 & -1\\ -1 &-1 & 1 \end{array}\right) \rightarrow \left(\begin{array}{ccc} \dfrac{1}{3} & 0.23 & -0.23\\ 0.23 & \dfrac{1}{3} & -0.23\\ -0.23 & -0.23 & \dfrac{1}{3} \end{array}\right),\nonumber\end{equation}

\begin{equation}\rho_{-,+}= \frac{1}{3}\left(\begin{array}{ccc} 1 & -1 & 1\\ -1 & 1 & -1\\ 1 &-1 & 1 \end{array}\right) \rightarrow \left(\begin{array}{ccc} \dfrac{1}{3} & -0.23 & 0.23\\ -0.23 & \dfrac{1}{3} & -0.23\\ 0.23 & -0.23 & \dfrac{1}{3} \end{array}\right),\nonumber\end{equation}

\begin{equation}\rho_{-,-}= \frac{1}{3}\left(\begin{array}{ccc} 1 & -1 & -1\\ -1 & 1 & 1\\ -1 &1 & 1 \end{array}\right) \rightarrow \left(\begin{array}{ccc} \dfrac{1}{3} & -0.23 & -0.23\\ -0.23 & \dfrac{1}{3} & 0.23\\ -0.23 & 0.23 & \dfrac{1}{3} \end{array}\right).\nonumber\end{equation}

Thus we see that the effect of the optimal cloning map on the $3$-pulse DPS states is similar to the depolarizing channel of the form 
\begin{equation}
    \operatorname{Tr}_{\rm Eve} [\Phi(\rho)]\rightarrow (1-p)\rho + \dfrac{p}{d}I,
\end{equation}
with $d=3$ and $p=0.31$. In other words, the set of states received by Bob after the optimal cloning attack can be written as,
\begin{equation}
  \tilde{\rho}_{\pm,\pm} =  \operatorname{Tr}_{\rm Eve}\Phi(\rho_{\pm,\pm}) \rightarrow (1-0.31)\rho_{\pm,\pm} + \dfrac{0.31}{3}I.
\end{equation}

\section{Effect of the Unitary cloner}\label{unitarycloner}
We describe the optimal unitary cloner and the best fidelity of cloning obtained for the $3$-pulse DPS states, following \cite{Mitra_2021}. Recall that the general symmetric cloning transformation described in Eq. \ref{cloningUnitary} 
\begin{equation}
\begin{aligned}
|e_i\rangle_A|0\rangle_B|X\rangle_C & \rightarrow p|e_i\rangle_A|e_i\rangle_B\left|X_i\right\rangle_C \\
& +q \sum_{j \neq i}~\left (|e_i\rangle_A|e_j\rangle_B+|e_j\rangle_A|e_i\rangle_B\right )~\left|X_j\right\rangle_C,
\end{aligned}
\end{equation}
is unitary when the parameters $p$ and $q$ satisfy $p^2 + 2(d-1)q^2 = 1$. For any ensemble of states, an analytical solution for the optimal values of $q$ and $p$ was derived in \cite{Mitra_2021}. In our case, the ensemble of states in consideration is the set of DPS states ${|\psi\rangle_{\pm.\pm}}$. Finding the best fidelity requires us to expand the ensemble in a basis such that the fourth power of the coefficients in the state expansion are maximized. This happens when the basis set is the eigenbasis for at least one of the states in the input ensemble. To this end, we consider the basis, already used in estimating the error rate for the optimal cloner in Sec.~\ref{sec:optimalcloning}.

\begin{eqnarray}
    |e_{1}\rangle &=&  \frac{1}{\sqrt{3}}\left[|1\rangle + |2\rangle + |3\rangle\right] \nonumber \\ |e_{2}\rangle &=& \frac{1}{\sqrt{6}}\left[|1\rangle + |2\rangle -2 |3\rangle\right] \nonumber \\ |e_{3}\rangle &=& \frac{1}{\sqrt{2}}\left[|1\rangle - |2\rangle\right] \nonumber.
\end{eqnarray} 
The DPS states can then be expanded in this basis as follows.
\begin{eqnarray}
|\psi\rangle_{+,+}&=&\frac{1}{\sqrt{3}}[|1\rangle +|2\rangle+|3\rangle] = |e_1\rangle \nonumber \\
|\psi\rangle_{+,-}&=&\frac{1}{\sqrt{3}}[|1\rangle +|2\rangle-|3\rangle] = \frac{1}{3}|e_1\rangle+\frac{2\sqrt{2}}{3}|e_2\rangle \nonumber\\
|\psi\rangle_{-,+}&=&\frac{1}{\sqrt{3}}[|1\rangle -|2\rangle+|3\rangle] \nonumber \\
&=& \frac{1}{3}|e_1\rangle-\frac{\sqrt{2}}{3}|e_2\rangle+\frac{\sqrt{2}}{\sqrt{3}}|e_3\rangle\nonumber \\
|\psi\rangle_{-,-}&=&\frac{1}{\sqrt{3}}[|1\rangle -|2\rangle-|3\rangle] \nonumber \\
&=& -\frac{1}{3}|e_1\rangle+\frac{\sqrt{2}}{3}|e_2\rangle+\frac{\sqrt{2}}{\sqrt{3}}|e_3\rangle .\nonumber
\end{eqnarray}
We find the optimal $q$ value using the ensemble of DPS states expanded in this basis, resulting in $q_{\rm opt} = 0.23$. In the qutrit basis $\{|1\rangle,|2\rangle,|3\rangle\}$), the transformation of the DPS states can be expressed as below.
\begin{equation}\rho_{+,+}= \frac{1}{3}\left(\begin{array}{ccc} 1 & 1 & 1\\ 1 & 1 & 1\\ 1 &1 & 1 \end{array}\right) \rightarrow \left(\begin{array}{ccc} \dfrac{1}{3} & 0.28 & 0.28\\ 0.28 & \dfrac{1}{3} & 0.28\\ 0.28 & 0.28 & \dfrac{1}{3} \end{array}\right),\nonumber\end{equation}
\begin{equation}\rho_{+,-}= \frac{1}{3}\left(\begin{array}{ccc} 1 & 1 & -1\\ 1 & 1 & -1\\ -1 &-1 & 1 \end{array}\right) \rightarrow \left(\begin{array}{ccc} 0.28 & 0.22 & -0.25\\ 0.22 & 0.28& -0.25\\ -0.25 & -0.25 & 0.44 \end{array}\right),\nonumber\end{equation}
\begin{equation}\rho_{-,+}= \frac{1}{3}\left(\begin{array}{ccc} 1 & -1 & 1\\ -1 & 1 & -1\\ 1 &-1 & 1 \end{array}\right) \rightarrow \left(\begin{array}{ccc} 0.36 & -0.25 & 0.14\\ -0.25 & 0.36& -0.17\\ 0.14 & -0.17 & 0.28 \end{array}\right),\nonumber\end{equation}
\begin{equation}\rho_{-,-}= \frac{1}{3}\left(\begin{array}{ccc} 1 & -1 & -1\\ -1 & 1 & 1\\ -1 &1 & 1 \end{array}\right) \rightarrow \left(\begin{array}{ccc} 0.36 & -0.25& -0.17\\ -0.25 & 0.36 & 0.14\\ -0.17 & 0.14 & 0.28 \end{array}\right).\nonumber\end{equation}
Similar to the analysis done in Sec.~\ref{sec:optimalcloning}, we find the error rate at Bob due to the action of the unitary cloner. $\rho_{+,+}$ after unitary cloning remains to be diagonal in the basis given above and we determine the action of the MZI on the basis and thus the state to find the probability of the erroneous click. For the rest of the states we do a spectral decomposition and find the action of MZI on the eigenbasis, which then can be used to find the error rate of the cloned states.

\end{appendices}
\bibliography{Ref}
\end{document}